\documentclass[vecphys]{svmult}

\usepackage{makeidx}         % allows index generation
\usepackage{graphicx}        % standard LaTeX graphics tool
\usepackage{multicol}        % used for the two-column index
\usepackage[bottom]{footmisc}% places footnotes at page bottom
\makeindex             % used for the subject index
\begin{document}

\title*{Large Magellanic Cloud Distance from Cepheid Variables using Least Squares Solutions}
% Use \titlerunning{Short Title} for an abbreviated version of
% your contribution title if the original one is too long
\author{C. Ngeow\inst{1}\and
S. M. Kanbur\inst{2}}
% Use \authorrunning{Short Title} for an abbreviated version of
% your contribution title if the original one is too long
\institute{University of Illinois, Urbana, IL 61801, USA
\texttt{cngeow@astro.uiuc.edu}
\and State University of New York at Oswego, Oswego, NY 13126, USA \texttt{kanbur@oswego.edu}}
%
% Use the package "url.sty" to avoid
% problems with special characters
% used in your e-mail or web address
%
\maketitle

%\begin{abstract} 
Distance to the Large Magellanic Cloud (LMC) is determined using the Cepheid variables in the LMC. We combine the individual LMC Cepheid distances obtained from the infrared surface brightness method and a dataset with a large number of LMC Cepheids. Using the standard least squares method, the LMC distance modulus can be found from the ZP offsets of these two samples. We have adopted both a linear P-L relation and a ``broken'' P-L relation in our calculations. The resulting LMC distance moduli are $18.48\pm0.03$ mag and $18.49\pm0.04$ mag (random error only), respectively, which are consistent to the adopted $18.50$ mag in the literature.
%\end{abstract}

\section{Introduction}
\label{sec:1}
% Always give a unique label
% and use \ref{<label>} for cross-references
% and \cite{<label>} for bibliographic references
% use \sectionmark{}
% to alter or adjust the section heading in the running head

\begin{figure}
\centering
\includegraphics[height=4cm]{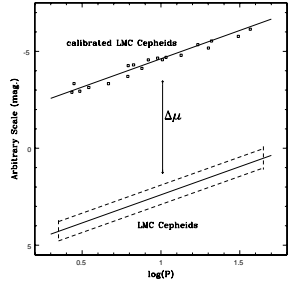}
\caption{Illustration of the LSQ. Introducing an offset, $\Delta \mu$, between the calibrated Cepheids and the other LMC Cepheids will allow us to simultaneously solve for the LMC distance modulus and the P-L relation with LSQ.}
\label{fig:1}      
\end{figure}

Recently, \cite{gie05} (hereafter G05) has used the infrared surface brightness method to obtain the individual distances to 13 LMC Cepheids with an averaged LMC distance modulus of $18.56\pm0.04$ mag (random error only). However, LMC hosts more than $600$ Cepheids with data available from the Optical Gravitational Lensing Experiment (OGLE) \cite{uda99}. A linear least squares solution (LSQ) will allow a simultaneous determination for {\it both} of the LMC distance and the P-L relation (see Figure \ref{fig:1}).

\section{Data, Method and Results}
\label{sec:2}

The data include the absolute magnitudes for the 13 Cepheids with individual distance measurements from G05 and the apparent magnitudes (after extinction correction) for $\sim630$ LMC Cepheids from \cite{kan06}, which is based on the OGLE database. For both datasets, we fit the following regression using the LSQ: $x=\alpha \Delta \mu + a + b \log(P)$ where $\alpha=0$ if $x=M$ (for G05 data) or $\alpha=1$ if $x=m$ (for OGLE data). The results of the LSQ are: $M^V=-2.76\pm0.04\log(P)-1.36\pm0.07$ with $\Delta \mu(V)=18.47\pm0.06$, and $M^I=-2.98\pm0.02\log(P)-1.86\pm0.05$ with $\Delta \mu(I)=18.48\pm0.04$. The weighted average of these $\Delta \mu$ is $18.48\pm0.03$ mag (random error only).

Since the recent studies have strongly suggested the LMC P-L relation is not linear (\cite{kan06,kan04,nge05,san04}), we also use the following ``broken'' regression to fit the data:

\begin{equation}
x  =  \alpha \Delta \mu + a_S + \beta b_S \log(P) + \gamma (a_L-a_S) + \epsilon b_L \log(P),
 \ \alpha = \left\{ \begin{array}{ll}
                      0 &  \mathrm{if}\ x = M \\
                      1 &  \mathrm{if}\ x = m
                     \end{array}
             \right . \nonumber
\end{equation}

\noindent where subscripts $_S$ and $_L$ refer to the short and long period Cepheids, respectively, and $\beta =1,\ \gamma=0,\ \epsilon=0,\ \mathrm{for}\ \log(P) < 1.0$; $\beta = 0,\ \gamma = 1,\ \epsilon = 1,\ \mathrm{for}\ \log(P) \geq 1.0$. The results are: $M^V_L=-2.84\pm0.16\log(P)-1.23\pm0.21;\ M^V_S=-2.94\pm0.06\log(P)-1.28\pm0.07$ with $\Delta \mu(V)=18.49\pm0.06$, and $M^I_L=-3.09\pm0.11\log(P)-1.69\pm0.14;\ M^I_S=-3.09\pm0.04\log(P)-1.80\pm0.05$ with $\Delta \mu(I)=18.49\pm0.04$. The weighted average of the distance moduli in both bands is $18.49\pm0.04$ mag (random error only). $F$-test (\cite{kan04,nge05}) is also applied to examine if the data is more consistent with a single-line regression (the null hypothesis) or a two-lines regression (the alternate hypothesis). For our data, we obtain $F(V)=7.3$ and $F(I)=6.9$, where $F\sim3$ at 95\% confident level. This suggested the null hypothesis can be rejected and the data is more consistent with the broken
P-L relation.

%
%
% BibTeX users please use
% \bibliographystyle{}
% \bibliography{}
%
% Non-BibTeX users please follow the syntax
% the syntax of "referenc.tex" for your own citations
%%%%%%%%%%%%%%%%%%%%%%%% referenc.tex %%%%%%%%%%%%%%%%%%%%%%%%%%%%%%
% sample references
% "physics"
%
% Use this file as a template for your own input.
%
%%%%%%%%%%%%%%%%%%%%%%%% Springer-Verlag %%%%%%%%%%%%%%%%%%%%%%%%%%

%
% BibTeX users please use
% \bibliographystyle{}
% \bibliography{}
%
% Non-BibTeX users please use

%%%%%%%%%%%%%%%%%%%%%%%%%%%%%%%%%%%%%%%%%%%%%%%%%%%%%%%%%%%%%%%%%%%%%%  }

%%%%%%%%%%%%%%%%%%%%%%%%%%%%%%%%%%%%%%%%%%%%%%%%%%%%%%%%%%%%%%%%%%%%%%

\printindex
\end{document}